%% file: main.tex
\documentclass[prd, 11pt,nofootinbib, superscriptaddress, preprintnumbers,floatfix]{revtex4}

\usepackage{iftex}

\ifPDFTeX
  \usepackage{ucs}
  \usepackage[utf8]{inputenc}
  \usepackage[T1]{fontenc}
  \usepackage[english]{babel}
\else
  \ifXeTeX
    \usepackage{xltxtra}
    \usepackage{polyglossia}
    \setmainlanguage[variant=british]{english}
    \setotherlanguage[spelling=new, latesthyphen=true]{german}
    \usepackage{fontspec}
    \defaultfontfeatures{Ligatures=TeX}
    \defaultfontfeatures{Mapping=tex-text}
  \else
    \usepackage{luatextra}
  \fi
\fi

\usepackage{amsmath,amssymb}
\usepackage{graphicx}
\usepackage{subfigure}
\usepackage{hyperref}

\def\Zq{Z_{\rm q}}

\begin{document}


\title{$\langle x\rangle$ and $\langle x^2\rangle$ of the pion PDF from
  Lattice QCD\\ with  
  $N_f=2+1+1$ dynamical quark flavours}

\date{\today}
\begin{center}
  \includegraphics[draft=false]{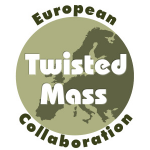}
\end{center}

\author{M.~Oehm}
\affiliation{Rheinische Friedrich-Wilhelms-Universität, Bonn, Germany}
\author{C.~Alexandrou}
\affiliation{University of Cyprus, Nicosia, Cyprus}
\affiliation{Computation-based Science and Technology Research Center, Nicosia, Cyprus}
\author{M.~Constantinou}
\affiliation{Temple University, Philadelphia, USA}
\author{K.~Jansen}
\affiliation{NIC/DESY Zeuthen, Germany}
\author{G.~Koutsou}
\affiliation{The Cyprus Institute, Nicosia, Cyprus}
\author{B.~Kostrzewa}
\affiliation{Rheinische Friedrich-Wilhelms-Universität, Bonn, Germany}
\author{F.~Steffens}
\affiliation{Rheinische Friedrich-Wilhelms-Universität, Bonn, Germany}
\author{C.~Urbach}
\affiliation{Rheinische Friedrich-Wilhelms-Universität, Bonn, Germany}
\email[Corresponding author: ]{urbach@hiskp.uni-bonn.de}
\author{S.~Zafeiropoulos}
\affiliation{Universität Heidelberg, Heidelberg, Germany}

\collaboration{\textbf{ETM Collaboration}}

\begin{abstract}
  Using $N_f{=}2{+}1{+}1$ lattice QCD, we determine the fermionic
  connected contributions to the first and second moment of the pion PDF. Based
  on gauge configurations from the European Twisted Mass
  Collaboration, chiral and continuum extrapolations are performed
  using pion masses in the range of 230 to 500 MeV and three values
  of the lattice spacing. Finite volume effects are investigated using
  different volumes. In order to avoid mixing under
  renormalisation for the second moment, we use an operator with
  two non-zero spatial components of momentum. Momenta are injected
  using twisted boundary conditions. Our final values read
  $\langle x\rangle_\mathrm{R}^\mathrm{phys}=
  0.2075(106)$ and $\langle x^2\rangle_\mathrm{R}^\mathrm{phys} = 0.163(33)$, determined at
  $2\ \mathrm{GeV}$ in the $\overline{\mathrm{MS}}$-scheme and with
  systematic and statistical uncertainties summed in quadrature. 
\end{abstract}

\maketitle


\input{intro}

\input{lattice}

\input{renormalisation}

\input{results}

\input{discussion}

\input{summary}

\bibliographystyle{h-physrev5}
\bibliography{bibliography,bib_pion}
\begin{appendix}
   \include{appendix}

\end{appendix}
\end{document}

%% file: intro.tex
\section{Introduction}
\label{sec:intro}

In Quantum Chromodynamics (QCD), the pion represents the Goldstone
boson of spontaneously broken chiral symmetry and is the lightest
hadronic state in the spectrum. As such it is of deep importance both
for the long range part of the nucleon-nucleon interaction  
and for the inner structure of the nucleon. In the latter case, it is now widely recognized 
that the pion is responsible for most, if not for all, of the excess of $\bar{d}$ over $\bar{u}$ 
anti-quarks in the proton sea~\cite{Thomas:2000ny,Chen:2001et,Chen:2001pva,Salamu:2014pka}.
Despite this importance, compared to the relatively detailed knowledge of the quark and gluon 
substructure of the nucleon, the pion sub-structure is largely unknown
because pion fixed target experiments cannot be built. Nevertheless,
Drell-Yan lepton-pair production and prompt photon production in  
totally inclusive pion-nucleon scattering \cite{Badier:1983mj,Betev:1985pf,Conway:1989fs}, 
as well as leading neutron electro production \cite{Aaron:2010ab} have been used to determine 
the pion structure functions.

Among the most important tools for understanding hadron structure are parton distribution functions 
(PDFs), which have been extensively studied both experimentally and theoretically. 
The determination of 
PDFs from experimental data requires fits based on phenomenological
models affected by systematic uncertainties that are not easy to
quantify. Therefore, a direct determination of parton distribution
functions from first principles is highly desirable. The method of
choice is thus lattice QCD, a non-perturbative tool based on
discretized Euclidean space-time. However, due to
their light-cone nature PDFs cannot 
be computed directly on a Euclidean lattice. Nevertheless, a recent proposal by
X.~Ji~\cite{Ji:2013dva} has led to the exciting possibility of
computing the Bjorken $x$ dependence of PDFs from lattice
QCD~\cite{Lin:2014zya,Alexandrou:2015rja,Chen:2016utp,Alexandrou:2016jqi}
based on quasi distributions instead of using the light cone. Indeed,
this method has been recently applied to the nucleon unpolarized
\cite{Alexandrou:2018pbm,Chen:2018xof}, helicity
\cite{Alexandrou:2018pbm,Lin:2018qky}, and transversity
\cite{Alexandrou:2018eet,Liu:2018hxv} distributions, directly at the physical
point, where the pion mass assumes its physical value. For the valence
quark distributions of the pion with mass of $M_\pi \approx 310$ MeV
results can be found in Ref.~\cite{Chen:2018fwa}. The alternative
proposal of using pseudo distributions was put forward by A.~Radyushkin in
Ref.~\cite{Radyushkin:2017cyf}. It was studied for the first time on the
lattice for the nucleon case in Ref.~\cite{Orginos:2017kos}. Its
relation to moments of PDFs is anlyzed in Ref.~\cite{Karpie:2018zaz}.
A third proposal by Y.-Q.~Ma and J.W.~Qiu can be found in
Ref.~\cite{Ma:2017pxb}.

Although these efforts have opened a new direction to access PDFs
demonstrating remarkable qualitative agreement with phenomenological
parametrizations, there are still a number of improvements that have
to be implemented before one reaches the reliability for a direct
quantitative comparison with systematic uncertainties under control.
For doubts on the aforementioned approaches we refer to
Ref.~\cite{Rossi:2018zkn}. 

In lattice QCD there is a long history of calculations of moments of
PDFs. In principle the PDFs can be obtained, as outlined in
Ref.~\cite{Capitani:1994qn} and references therein, using the inverse
Mellin transform and the operator product expansion. 
Such a reconstruction can only be reliable if several moments of PDFs are 
available~\cite{Detmold:2001dv}. However, the signal-to-noise ratio decreases for high moments and mixing with 
lower dimensional operators becomes unavoidable. Nevertheless, there have been advances in noise reduction techniques 
and methods to disentangle mixing between operators, which allow one to extract moments beyond the leading one. This 
progress has led to investigations of interesting physics questions, such as the momentum and spin decomposition of 
the nucleon in terms of their quark and gluon contents. Within our ETM
collaboration, this has been accomplished by lattice QCD simulations 
directly at the physical point~\cite{Alexandrou:2017oeh}, where both momentum and 
spin sum rules have been verified without imposing any
constraints. For the pion, however, the situation is much less
satisfactory. Earlier studies have computed the first three moments either in the quenched approximation 
\cite{Best:1997qp,Guagnelli:2004ga,Capitani:2005jp} or for connected insertions only \cite{Brommel:2006zz,Baron:2007ti,Bali:2013gya,Abdel-Rehim:2015owa}, all of them using simulations with quark masses away from their physical value. Only a few results for realistic QCD
simulations appear in the literature, that is,
Ref.~\cite{Bali:2013gya} at about 150 MeV for the pion mass, and a
determination directly at the physical point with $N_f=2$ in
Ref.~\cite{Abdel-Rehim:2015owa} for the lowest moment. Given the importance of the pion 
for ongoing and planned experiments, further study of the pion structure is imperative. For the extraction of reliable estimates, 
systematic uncertainties such as discretisation and volume effects must be properly addressed and quantified.

In this work we present a multi-component effort in the aforementioned direction, with a variety of improvements 
compared to the studies available in the literature, in terms of the
ensembles employed and level of control over systematic uncertainties
in the computed moments. We 
calculate the light quark connected contributions of the first and
second moment -- $\langle x\rangle$ and $\langle x^2\rangle$ --
of the pion using lattice QCD simulations that include degenerate
light as well as strange and charm quarks in the
sea ($N_f{=}2{+}1{+}1$). We use several ensembles produced by the European Twisted Mass Collaboration (ETMC) corresponding 
to three values of the lattice spacing, which allow us to study discretisation effects. These ensembles have pion mass 
values that range between 230 MeV and 500 MeV, which are combined in a chiral extrapolation to obtain the value at the
physical pion mass. Different volumes are used to investigate finite
size effects and excited state contaminations.
In addition, a way around possible mixing for $\langle x^2\rangle$ is the choice of an operator that 
is free from mixing under renormalisation. A first preliminary account of this
work can be found in Ref.~\cite{Kostrzewa:2016vbz}.

The remainder of the paper is organized as follows: In Section~\ref{sec:lattice} we discuss the technical aspects of the 
lattice calculation, while in Section~\ref{sec:renormalisation} we discuss the method used for the determination of the
required renormalisation functions in the RI'-MOM scheme and the
conversion to the $\overline{\mathrm{MS}}$-scheme. The main results 
of this work are presented in Section~\ref{sec:results}, followed by a
discussion and a summary in Section~\ref{sec:discussion}. Technical
details related to renormalisation can be found in Appendix~\ref{appA}, while
correlation coefficients of fit parameters are collected in Appendix~\ref{sec:appB}.

%% file: lattice.tex
\section{Lattice Details}
\label{sec:lattice}

The calculation presented in this paper is based on gauge configurations generated
by ETMC with $N_f=2+1+1$ dynamical quark flavors at three values of the 
lattice spacing. Details for the configuration generation and analyses for basic
quantities can be found in
Refs.~\cite{Chiarappa:2006ae,Baron:2010th,Baron:2010bv}. The ensembles
were generated using the Iwasaki gauge action~\cite{Iwasaki:1985we}
and the Wilson twisted mass fermion action at maximal
twist~\cite{Frezzotti:2003ni,Frezzotti:2003xj,Frezzotti:2004wz}. Working
at maximal twist 
guarantees $\mathcal{O}(a)$ improvement for most physical
quantities~\cite{Frezzotti:2003ni}, and in particular for the
quantities considered here. 

\begin{table}[t!]
  \centering
  \begin{tabular*}{.9\textwidth}{@{\extracolsep{\fill}}lccccccc}
    \hline\hline
    ensemble & $\beta$ & $a\mu_\ell$ & $a\mu_\sigma$ & $a\mu_\delta$ &
    $L/a$ & $T/a$ & $N_\mathrm{conf}$  \\ 
    \hline\hline
    A30.32   & $1.90$ & $0.0030$ & $0.150$  & $0.190$  & $32$ & $64$ & $280$ \\
    A40.24   & $1.90$ & $0.0040$ & $0.150$  & $0.190$  & $24$ & $48$ & $280$ \\
    A40.32   & $1.90$ & $0.0040$ & $0.150$  & $0.190$  & $32$ & $64$ & $250$ \\
    A60.24   & $1.90$ & $0.0060$ & $0.150$  & $0.190$  & $24$ & $48$ & $313$ \\
    A80.24   & $1.90$ & $0.0080$ & $0.150$  & $0.190$  & $24$ & $48$ & $304$ \\
    A100.24  & $1.90$ & $0.0100$ & $0.150$  & $0.190$  & $24$ & $84$ & $312$ \\

    \hline
    B25.32   & $1.95$ & $0.0025$ & $0.135$  & $0.170$  & $32$ & $64$ & $212$ \\
    B35.32   & $1.95$ & $0.0035$ & $0.135$  & $0.170$  & $32$ & $64$ & $249$ \\

    B55.32   & $1.95$ & $0.0055$ & $0.135$  & $0.170$  & $32$ & $64$ & $310$ \\
    B85.24   & $1.95$ & $0.0085$ & $0.135$  & $0.170$  & $24$ & $84$ & $357$ \\
    \hline
    D15.48   & $2.10$ & $0.0015$ & $0.120$  & $0.1385$ & $48$ & $96$ & $161$ \\
    D30.48   & $2.10$ & $0.0030$ & $0.120$  & $0.1385$ & $48$ & $96$ & $174$ \\
    D45.32sc & $2.10$ & $0.0045$ & $0.0937$ & $0.1077$ & $32$ & $64$ & $300$ \\
    \hline\hline
    \vspace*{0.1cm}
  \end{tabular*}
  \caption{The $N_f=2+1+1$ ensembles used in this investigation.
    The notation of Ref.~\cite{Baron:2010bv} is used for labeling the
    ensembles. We list the bare parameters $\beta$, $\mu_\ell$,
    $\mu_\sigma$ and $\mu_\delta$. $T/a$ and $L/a$ are time and
    spatial extents of the lattice, respectively. $N_\mathrm{conf}$ is
    the number of configurations we used to estimate the moments.}
  \label{tab:setup211}
\end{table}

The bare parameters of the ensembles used here are summarized in
Table~\ref{tab:setup211}. $\mu_\ell$ is the bare light quark mass
directly proportional to the renormalised light quark
mass. $\mu_\sigma$ and $\mu_\delta$ parametrize the strange and charm
quark masses~\cite{Frezzotti:2003xj,Chiarappa:2006ae}.
For the subset of configurations we used from each ensemble we have
computed the autocorrelation times for the relevant quantities to
verify their statistical independence.
The error analysis is performed using the stationary blocked bootstrap
procedure~\cite{Dimitris:1994} with $1500$ bootstrap samples.

In general, the computation of the moments requires the computation of
three-point functions of the form
\begin{equation}
  \label{eq:3pt}
  C_{\mathcal{O}}(t, \vec p)=\sum_{\vec x,\vec y}\langle \pi(T/2,\vec
  x, \vec p)\,
  \mathcal{O}(t,\vec y)\,
  \pi^\dagger(0,\vec0, \vec p)\rangle
\end{equation}
with operator $\mathcal{O}$ inserted at Euclidean time $t$. We fix
here the time difference between the two pions to $T/2$, which is not
necessary, but convenient. The operators for the two moments will be
detailed below. The particular choice of operators is motivated
by their transformation properties under the symmetries of the lattice
as well as the requirement of minimal mixing with lower-dimensional
operators under renormalisation, see Refs.\cite{Capitani:1994qn,Beccarini:1995iv,Gockeler:1996mu}
for details. The interpolating operators for the pions read
\begin{equation}
  \pi(t, \vec x, \vec p)\ =\ \bar\psi(t, \vec x, \vec\theta)\,
  i\gamma_5\ \frac{\tau^1 + i\tau^2}{2}\ \psi(t, \vec x, \vec\theta^\prime)\,,
\end{equation}
with the momentum $\vec p\propto\vec\theta-\vec\theta^\prime$ realized
via twisted boundary conditions, see 
below. $\tau^i, i=1,2,3$ are the Pauli matrices acting in flavor 
space and $\psi=(u, d)^t$ is the light quark field.

\subsection{The first moment $\langle x\rangle$}

A convenient operator in Euclidean space-time for the calculation of
the first moment $\langle x\rangle$ is 
\begin{equation}
  \label{eq:O44}
  \mathcal{O}_{v2b} \equiv\mathcal{O}_{44}(x)=\frac{1}{2}\bar \psi(x)[\gamma_4
  \stackrel{\leftrightarrow}{D}_4-\frac{1}{3}\sum_{k=1}^{3}\gamma_k
  \stackrel{\leftrightarrow}{D}_k]\left(\frac{1+\tau^3}{2}\right)\psi(x)\,.
\end{equation}
Here, $\stackrel{\leftrightarrow}{D}_\mu=\frac{1}{2} 
(\bigtriangledown_\mu+\bigtriangledown_\mu^*)$ is the symmetric,
gauge covariant lattice derivative with 
$\bigtriangledown_\mu$ ($\bigtriangledown_\mu^*$) being the usual 
gauge covariant forward (backward) derivative on the lattice.
The above operator has the advantage that $\langle x\rangle$ is extracted without 
the need for an external momentum, because external momentum in general
increases the noise. For the first use of this operator with Wilson
twisted mass fermions we refer to Refs.~\cite{Capitani:2005aa,Capitani:2005jp}. 

\begin{figure}[t]
  \centering
  \includegraphics[width=.7\linewidth]{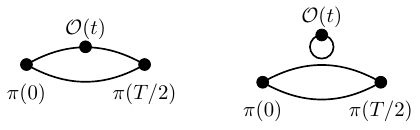}
  \caption{Connected (left) and disconnected (right) contributions to
    the three-point functions. The lines
    represent quark propagators.}
  \label{fig:diagrams}
\end{figure}

The bare moment $\langle x\rangle_\mathrm{bare}$ is related to the
matrix element of the operator $\mathcal{O}_{44}$ as follows
\begin{equation}
  \langle\pi(p)| \mathcal{O}_{44}|\pi(p)\rangle\ =\ 2 (p^0p^0 -
  \frac{1}{3}\vec p\, \vec p\,)\langle x\rangle_\mathrm{bare}\,,
\end{equation}
where $p=(p^0, \vec{p})$ is the four momentum of the pions.
With pions at rest one obtains
\begin{equation}
  \langle x\rangle_\mathrm{bare}\ =\
  \frac{1}{2M_\pi^2}\langle\pi(0)|\mathcal{O}_{44}
  |\pi(0)\rangle,
\end{equation}
with $M_\pi$ the mass of the pion.
The matrix element
$\langle\pi(0)|\mathcal{O}_{44}|\pi(0)\rangle$ between two
pions at rest is calculated from the ratio 
\begin{equation}
  \langle\pi(0)|\mathcal{O}_{44}|\pi(0)\rangle=
  4M_\pi\frac{C_{44}(t, \vec 0)}{C_\pi(T/2, \vec 0)}\quad
  (0\ll t\ll T/2)
  \label{eq:averx_ratio}
\end{equation}
of the three point function
\begin{equation}
  \label{eq:C44}
  C_{44}(t, \vec 0)=\sum_{\vec x,\vec y}\langle \pi(T/2,\vec x, \vec 0)\,
  \mathcal{O}_{44}(t,\vec y, \vec 0)\,
  \pi^\dagger(0,\vec0, \vec0)\rangle\,,
\end{equation}
over the two point function
\begin{equation}
  C_\pi(T/2, \vec p)=\sum_{\vec x}\langle \pi(T/2,\vec x, \vec
  p)\pi^\dagger(0,\vec 0, \vec p)\rangle\,.
\end{equation}
In Eq.~\ref{eq:averx_ratio} a factor of $2M_\pi$ relates the lattice and continuum matrix
elements of $\mathcal{O}_{44}$ between pion states and a further factor of $2$ relates
the ratio of correlation functions to the value of the matrix element.
This leads to
\begin{equation}
  \label{eq:xbare}
  \langle x\rangle_\mathrm{bare}(t)\ =\ \frac{2}{M_\pi}
  \frac{C_{44}(t, \vec 0)}{C_\pi(T/2, \vec 0)}\quad  (0\ll t\ll T/2)\,. 
\end{equation}
There are two contributions in the Wick contractions of $C_{44}$: the first is
extracted when the current couples to the quarks of the pion directly (connected diagram),
while the second is obtained from the so-called quark loop (disconnected diagram) in which
the current interacts with the pion via gluon exchange. Both are
visualized in Fig.~\ref{fig:diagrams}. The disconnected contribution is ignored 
in our calculation, assuming that it is small, which is indeed the
case for the nucleon~\cite{Abdel-Rehim:2013wlz}. A computation of the
disconnected contributions is, however, planned for the near future.

\subsection{The second moment $\langle x^2\rangle$}

In order to avoid mixing under renormalisation with lower dimensional
operators~\cite{Beccarini:1995iv}, we use for the second moment the
operator (in Euclidean space-time) 
\begin{equation}
  \label{eq:O012}
  \mathcal{O}_{012}\ =\ \frac{1}{2^2}\bar\psi\gamma_{\{0}
  \stackrel{\leftrightarrow}{D}_1 \stackrel{\leftrightarrow}{D}_{2\}}
  \left(\frac{1+\tau^3}{2}\right)\psi - \mathrm{traces}\,.
\end{equation}
which is related to $\langle x^2\rangle$ via
\begin{equation}
  \langle \pi(p)| \mathcal{O}_{012} | \pi(p)\rangle\ =\ -2(p^0 p^1 p^2)
  \langle x^2\rangle_\mathrm{bare}\,.
\end{equation}
In contrast to $\langle x\rangle$, non-zero momentum is needed to
extract $\langle x^2\rangle$, due to the presence of the kinematic factor 
$(p^0 p^1 p^2)$ multiplying the quantity of interest. We use twisted boundary
conditions here to inject momentum, see below. As in the case of the first moment, 
the matrix element of the second moment is related to a 
ratio of three point to two point function
\begin{equation}
  \langle \pi(p)| \mathcal{O}_{012} |
  \pi(p)\rangle\ =\ 4 E_\pi(\vec p) \frac{C_{012}(t, \vec p)}{C_\pi(T/2, \vec p)}\,,
\end{equation}
which leads to
\begin{equation}
  \label{eq:x2bare}
  \langle x^2\rangle_\mathrm{bare}(t)\ =\ \frac{2}{p^1
    p^2}\ \frac{C_{012}(t, \vec p)}{C_\pi(T/2, \vec p)}\,. 
\end{equation}
For details on the implementation of $\mathcal{O}_{012}$ we refer to 
Ref.~\cite{Beccarini:1995iv}. We employ the convention given therein
for the discretisation of terms involving $\stackrel{\rightarrow}{D}
\stackrel{\leftarrow}{D}$. 

\subsection{The pion mass and decay constant}

\begin{table}[t!]
  \centering
  \begin{tabular*}{.9\textwidth}{@{\extracolsep{\fill}}lcccccc}
    \hline\hline
    ensemble & $\theta$ & $M_\pi$ & $f_\pi$ & $K_f$  & $K_m$ & $M_\pi L$\\
    \hline
    A30.32 & 0.4242640 & \(0.12361(48)\) & \(0.06459(25)\) & 0.9757 & 1.0023
    & \(4.0\)\\
    A40.24 & 0.2828425 & \(0.14423(62)\) & \(0.06567(34)\) & 0.9406 & 1.0099
    & \(3.5\)\\
    A40.32 & 0.3771235 & \(0.14147(47)\) & \(0.06809(22)\) & 0.9874 & 1.0013
    & \(4.5\)\\
    A60.24 & 0.3535535 & \(0.17253(72)\) & \(0.07148(26)\) & 0.9716 & 1.0047
    & \(4.1\)\\
    A80.24 & 0.3535535 & \(0.19953(48)\) & \(0.07596(21)\) & 0.9839 & 1.0025
    & \(4.8\)\\
    A100.24 & 0.4242640 & \(0.22117(49)\) & \(0.07931(22)\) & 0.9900 &
    1.0015 & \(5.3\)\\
    \hline
    B25.32 & 0.4242640 & \(0.10882(52)\) & \(0.05518(32)\) & 0.9605 &
    1.0136 & \(3.5\)\\
    B35.32 & 0.4242640 & \(0.12450(53)\) & \(0.06056(22)\) & 0.9794 & 1.0025
    & \(4.0\)\\
    B55.32 & 0.4242640 & \(0.15534(28)\) & \(0.06513(16)\) & 0.9920 & 1.0009
    & \(5.0\)\\
    B85.24 & 0.4242640 & \(0.19253(53)\) & \(0.06984(21)\) & 0.9795 & 1.0032
    & \(4.6\)\\
    \hline
    D15.48 & 0.5185450 & \(0.06986(43)\) & \(0.04298(20)\) & 0.9762 & 1.0081
    & \(3.4\)\\
    D30.48 & 0.4714045 & \(0.09786(28)\) & \(0.04721(13)\) & 0.9938 & 1.0021
    & \(4.7\)\\
    D45.32 & 0.3771235 & \(0.11980(48)\) & \(0.04826(18)\) & 0.9860 &
    1.0047 & \(3.8\)\\
    \hline\hline
  \end{tabular*}
  \caption{$\theta$-values, pion mass $aM_\pi$, pion decay constant
    $af_\pi$, finite volume correction factors $K_f$ and $K_m$ as well
    as $M_\pi L$ for all ensembles.}
  \label{tab:thetas}
\end{table}

The pion mass enters the equations leading to $\langle x\rangle$ and
$\langle x^2\rangle$, and thus, it must be computed. It can be obtained 
by fits of the functional
form
\begin{equation}
  \label{eq:fit2pt}
  f(t, A, M_\pi)\ =\ A \left(e^{-M_\pi t} + e^{-M_\pi(T-t)}\right)
\end{equation}
to the data for $C_\pi(t)$ for sufficiently large Euclidean times.
In twisted mass lattice QCD at maximal twist the pion decay constant
is directly related to the amplitude $A$ via
\begin{equation}
  f_\pi = 2\mu_\ell \frac{\sqrt{A}}{M_\pi^3}
\end{equation}
without the need for renormalisation~\cite{Jansen:2003ir}.

We note that $M_\pi$ and $f_\pi$ are affected significantly by finite size
effects~\cite{Carrasco:2014cwa}. Therefore, we use the corrections computed in
Ref.~\cite{Carrasco:2014cwa}, which are summarized in table~\ref{tab:thetas} 
for all the ensembles used in this work. 

In Eqs.~\ref{eq:xbare} and \ref{eq:x2bare} one needs to divide by
the two-point function at $T/2$. We explore two possibilities to
perform this division: the first one is to use the data of $C_\pi$ for
$t=T/2$. The second one is to first fit Eq.~\ref{eq:fit2pt} to the
data for $C_\pi$ in the region, where the ground state dominates and
then use the best fit parameters to reconstruct $C_\pi(T/2)$. The
latter procedure can help to average out fluctuations. However, the
differences between the two procedures are always 
well below the statistical uncertainty both for $\langle x\rangle$ and
$\langle x^2\rangle$. As our method of choice, we henceforth use 
the second of the two methods. 

\subsection{Stochastic Evaluation}

The above two and three point correlators are evaluated by using a stochastic
time slice source (Z(2)-noise in both real and imaginary part)
\cite{Dong:1993pk,Foster:1998vw,McNeile:2006bz} for all
color, spin and spatial indices. i.e., the quark propagator
$X^{b\vec\theta}_\beta(y)$ for quark flavour $f$ and twist angle $\vec\theta$ is
obtained by solving\footnote{Greek indices represent spin and latin
  indices color degrees of freedom. $f=u,d$ indexes the (light) quark flavours.}
\begin{equation}
  \sum_{y, \beta, b} D^{ab\vec\theta}_{\alpha\beta f}(z,y)X^{b\vec\theta}_{\beta f}(y)=\xi(\vec z)^a_\alpha
  \delta_{z_0,0}\quad (\mbox{source at}~ t=0)
\end{equation}
for $X$,
where the Z(2) random source $\xi(\vec z)^a_\alpha$ satisfies
the random average condition
\begin{equation}
  \langle\xi^*(\vec x)^a_\alpha\xi(\vec y)^b_\beta\rangle=
  \delta_{\vec x,\vec y}\delta_{a,b}\delta_{\alpha,\beta}\,.
\end{equation}
This allows one to estimate for instance the pion two-point function
$C_\pi(t)$ at zero momentum from
\[
C_\pi(t, \vec 0)\ =\ \sum_{\vec x, a, \alpha} X^{a\vec 0}_{\alpha f}(\vec x,
t)\,\cdot [X^{a \vec 0}_{\alpha f}(\vec x, t)]^*\ +\ \textrm{noise}\,,
\]
where the $\gamma_5$-hermiticity $D_u = \gamma_5 D_d^\dagger \gamma_5$
has been used. 
The generalized propagator\cite{Martinelli:1987zd} $\Sigma^{b\vec\theta'\vec\theta}_{\beta f' f}(y)$
needed in the computation of
$C_{44}(t)$ is obtained by solving
\begin{equation}
  \sum_yD^{ab\vec\theta'}_{\alpha\beta f'}(z,y)\ \Sigma^{b\vec\theta'\vec\theta}_{\beta f'
    f}(y)=\gamma_5X^{a\vec\theta}_{\alpha f}(z)\
  \delta_{z_0,T/2}\quad (\mbox{sink at}~ t=T/2)
\end{equation}
for $\Sigma$.
This approach was first applied for $\langle x\rangle$ of the pion in
Ref.~\cite{Baron:2007ti} and we used it recently in a computation of
the pion vector form
factor~\cite{Kostrzewa:2016vbz,Alexandrou:2017blh}, where further
details can be found. 
To further improve the signal, we use $N_\mathrm{src}=5$ sources per
gauge configuration and average. The source time slices are chosen
uniformly random in $[0, 1, \ldots, T-1]$.

\subsection{Twisted Boundary Conditions}

In order to realize non-zero momentum of arbitrary values for the
pions as needed for $\langle x^2\rangle$, we make use
of so-called twisted boundary
conditions~\cite{Guagnelli:2003hw,Bedaque:2004kc,deDivitiis:2004kq}. Enforcing
the spatial boundary conditions $\psi(x + \vec e_i L) = e^{2\pi \mathrm{i}\theta_i}\psi(x)$
on the quark fields changes the momentum quantization condition in
finite volume to $p_i = \frac{2\pi\theta_i}{L} + \frac{2\pi n_i}{L}$.
In time direction we chose $\theta_0=1/2$ to obtain anti periodic
boundary conditions in time. We chose the $\vec\theta$ in the spatial
directions non-zero to obtain non-zero momentum for the pions.

For the two quarks in the pion, we always chose zero twist angle for one
of the quarks and non-zero $\vec{\theta}$ for the other one. The pion
three-momentum $\vec p$ is then given by ($n_i = 0$)
\[
\vec{p}\ =\ \frac{2\pi\vec{\theta}}{L}\,.
\]
We recall that for the computation of
$\langle x^2\rangle$ two non-zero spatial components of the pion
momentum are needed when $\mathcal{O}_{012}$ is used, see
Eq.~\ref{eq:x2bare}. 
We chose the two non-zero elements of $\vec{\theta}$ equal,
i.e. for instance $\vec{\theta}=(\theta, \theta, 0)$. The
corresponding values for $\theta$ for each ensemble are compiled in
table~\ref{tab:thetas}. 
We always perform the computation for $\langle x^2\rangle$ for both
$\pm \vec p$ and average. The such obtained result is automatically
$\mathcal{O}(a)$ improved.

The main reason for using twisted boundary conditions is the fact that
noise in the three point and two point functions increases
significantly with increasing modulus of the injected momentum. With
twisted boundary conditions we are able to chose the momentum as small
as possible. However, we remark that twisted boundary conditions
induce additional finite 
volume effects, which might influence our
results~\cite{Jiang:2006gna}. As will be 
discussed later, we do not see such effects in $\langle x^2\rangle$
within statistical uncertainties.

\subsection{Chiral Extrapolations}

In Ref.~\cite{Diehl:2005rn} the pion mass dependence of pion moments
has been computed in leading order (LO) chiral perturbation theory
(ChPT). The functional form for $\langle x\rangle$ reads
\begin{equation}
  \label{eq:ChPTx}
  \langle x\rangle (M_\pi^2)\ =\ c_0 + c_1 \frac{M_\pi^2}{f_\pi^2}
\end{equation}
with low energy constants (LECs) $c_0$ and $c_1$. For the second moment
it reads
\begin{equation}
  \label{eq:ChPTx2}
  \langle x^2\rangle (M_\pi^2)\ =\ b_0\left(1 - \frac{M_\pi^2}{(4\pi
f_\pi)^2}\log\frac{M_\pi^2}{\mu_R^2}\right) + b_1 \frac{M_\pi^2}{f_\pi^2}\,,
\end{equation}
where we denote the corresponding LECs with $b_0$ and $b_1$. We chose
the renormalisation scale conventionally $\mu_R = f_\pi$. In contrast
to Ref.~\cite{Diehl:2005rn}, we have expressed the two moments as a
function 
of $M_\pi/f_\pi$, which has the big advantage of fully dimensionless
expressions. In principle one should then use $f_\pi^\mathrm{phys}$,
i.e. the physical value of the decay constant. However, we use here
$f_\pi$ as estimated for each ensemble, 
because scale setting is required only to estimate the moments at the
physical point. Since $f_\pi$ is a constant in leading order ChPT,
this procedure is consistent to the order of ChPT we are working here.
Unfortunately, the next-to-leading-order expressions for the moments
are not known. Still, in contrast to the case of nucleons, in the pion
sector ChPT works well such that we expect already the lowest order
to provide a reliable tool for our set of pion masses.
In order to account also for lattice spacing artifacts
we add terms $c_a a^2/r_0^2$ and $b_a a^2/r_0^2$ to the expressions
for the first and the second moment, respectively.

%% file: renormalisation.tex
\section{Renormalisation Functions}
\label{sec:renormalisation}

A renormalisation factor (Z-factor) must be applied to the bare matrix
elements of the operators defined in Eq.~(\ref{eq:O44}) and
Eq.~(\ref{eq:O012}), in order to obtain physical quantities.
More precisely, the bare and the renormalised moments are related as
follows 
\begin{equation}
  \langle x\rangle_\mathrm{R} = Z_\mathrm{vD}\langle
  x\rangle_\mathrm{bare}\,,\qquad \langle x^2\rangle_\mathrm{R} = Z_\mathrm{vDD}\langle x^2\rangle_\mathrm{bare}\,.
\end{equation}
In particular, the renormalisation procedure eliminates divergences with
respect to the lattice regulator, and allows the continuum limit to be
taken. In this Section we present the methodology and results for the
renormalisation functions, which are finally converted to the
$\overline{\rm MS}$-scheme at a scale $\overline{\mu}{=}2\ \mathrm{GeV}$. 
We employ the Rome-Southampton method (RI$'$ scheme)~\cite{Martinelli:1994ty} 
to compute the Z-factors non-perturbatively determined by the conditions 
\begin{eqnarray}
   \Zq = \frac{1}{12} {\rm Tr} \left[(S^L(p))^{-1}\, S^{{\rm Born}}(p)\right] \Bigr|_{p^2=\mu_0^2}\,,  \label{Zq_cond}\\[2ex]
   \Zq^{-1}\,Z_{\cal O}\,\frac{1}{12} {\rm Tr} \left[\Gamma^L(p)
     \,\Gamma^{{\rm Born}-1}(p)\right] \Bigr|_{p^2=\mu_0^2} &=& 1\, .
\label{renormalization cond}
\end{eqnarray}
The momentum $p$ is set to the RI$'$ renormalisation scale, $\mu_0$, $S^{{\rm Born}}$ 
($\Gamma^{{\rm Born}}$) is the tree-level value of the fermion propagator (operator), 
and the trace is taken over spin and color indices. 

We obtain the Z-factors using several ensembles at different values of
the pion mass, so that the chiral limit can be safely taken. In
addition, on each ensemble we use several values of the momentum $p$
(to be set equal to the RI$'$ renormalisation scale $\mu_0$) to
control systematic uncertainties as explained below. The RI$'$ values
for the Z-factors are converted to the ${\overline{\rm MS}}$ scheme
and are evolved to a reference scale of $2\ \mathrm{GeV}$ using an 
intermediate Renormalisation Group Invariant scheme defined in continuum 
perturbation theory. Renormalized matrix elements can be compared to 
phenomenological and experimental estimates that typically refer to 
quantities renormalized in the ${\overline{\rm MS}}$ scheme.

For a proper chiral extrapolation we compute the Z-factors on ensembles 
generated specifically for the renormalisation program of ETMC that include 
four degenerate quarks ($N_f{=}4$) at the same values of $\beta$ as the 
$N_f{=}2{+}1{+}1$ ensembles used for the calculation of $\langle x \rangle$ 
and $\langle x^2 \rangle$. The parameters of the ensembles are given in 
Table~\ref{tab:ensembles}, where the lattice spacing is determined using 
the nucleon mass computed with the $N_f{=}2{+}1{+}1$ twisted mass 
configurations~\cite{Alexandrou:2013joa, Abdel-Rehim:2015jna}.

\begin{table}[!h]
  \begin{center}
    \begin{tabular}{ccccc}
      \hline
      \hline
      $a \mu$ & $\kappa$ & $a \mu^{sea}_{\rm PCAC}$ & $a M_\mathrm{PS}$ & lattice size\\
      \hline
      \hline
      $\,\,\,$        $\,\,\,$   &             & $\beta=1.90$, $a=0.0934$ fm  &        &           \\ \hline
      $\,\,\,$  0.0080$\,\,\,$   &  0.162689   & $+$0.0275(4)     & 0.280(1)   & $24^3 \times 48$ \\  
      $\,\,\,$        $\,\,\,$   &  0.163476   & $-$0.0273(2)     & 0.227(1)   \\ \hline              
      $\,\,\,$  0.0080$\,\,\,$   &  0.162876   & $+$0.0398(1)     & 0.279(2)   & $24^3 \times 48$ \\  
      $\,\,\,$        $\,\,\,$   &  0.163206   & $-$0.0390(1)     & 0.241(1)   \\ \hline              
      $\,\,\,$        $\,\,\,$   &             & $\beta=1.95$, $a=0.082$ fm  &         &          \\ \hline
      $\,\,\,$  0.0020$\,\,\,$   &  0.160524   & $+$0.0363(1)     &    & $24^3 \times 48$ \\          
      $\,\,\,$        $\,\,\,$   &  0.161585   & $-$0.0363(1)     &    \\ \hline                      
      $\,\,\,$  0.0085$\,\,\,$   &  0.160826   & $+$0.0191(2)     & 0.277(2)   & $24^3 \times 48$ \\  
      $\,\,\,$        $\,\,\,$   &  0.161229   & $-$0.0209(2)     & 0.259(1)   \\ \hline              
      $\,\,\,$  0.0180$\,\,\,$   &  0.160826   & $+$0.0163(2)     & 0.317(1)   & $24^3 \times 48$ \\  
      $\,\,\,$        $\,\,\,$   &  0.161229   & $-$0.0160(2)     & 0.292(1)   \\ \hline              
      $\,\,\,$        $\,\,\,$   &             & $\beta=2.10$, $a=0.064$ fm    &      &           \\ \hline
      $\,\,\,$  0.0030$\,\,\,$   &  0.156042   & $+$0.0042(1)     & 0.127(2)   & $32^3 \times 64$ \\  
      $\,\,\,$        $\,\,\,$   &  0.156157   & $-$0.0040(1)     & 0.129(3)   \\ \hline              
      $\,\,\,$  0.0046$\,\,\,$   &  0.156017   & $+$0.0056(1)     & 0.150(2)   & $32^3 \times 64$ \\  
      $\,\,\,$        $\,\,\,$   &  0.156209   & $-$0.0059(1)     & 0.160(4)   \\ \hline              
      $\,\,\,$  0.0064$\,\,\,$   &  0.155983   & $+$0.0069(1)     & 0.171(1)   & $32^3 \times 64$ \\  
      $\,\,\,$        $\,\,\,$   &  0.156250   & $-$0.0068(1)     & 0.180(4)   \\ \hline              
      $\,\,\,$  0.0078$\,\,\,$   &  0.155949   & $+$0.0082(1)     & 0.188(1)   & $32^3 \times 64$ \\  
      $\,\,\,$        $\,\,\,$   &  0.156291   & $-$0.0082(1)     & 0.191(3) \\                       
      \hline
      \hline
    \end{tabular}
    \caption{Simulation details for the ensembles used for the renormalisation functions.} 
    \label{tab:ensembles}
  \end{center}
\end{table}

We employ the momentum source method introduced in Ref.~\cite{Gockeler:1998ye} 
and used in Ref.~\cite{Alexandrou:2015sea}, which leads to a high statistical 
accuracy with a small number of configurations. For the Z-factors presented in 
this work we use between 10 to 50 configurations depending on the ensemble under 
study. To reduce discretisation effects we use momenta that have the same spatial 
components, that is:
\begin{equation}
(a\,p) \equiv 2\pi \left(\frac{n_t}{L_t}+\frac{1}{2\,L_t},
\frac{n_x}{L_s},\frac{n_x}{L_s},\frac{n_x}{L_s}\right)\,,  \qquad\quad n_t \,\epsilon\, 
[2, 20]\,,\quad n_x\,\epsilon\, [1, 10]\,,
\end{equation} 
where $L_t$ ($L_s$) is the temporal (spatial) extent of the lattice, 
and we restrict the momenta up to $(a\,p)^2{\sim} 7$. A useful constraint 
for the chosen spatial momenta is $ {\sum_i p_i^4}/{(\sum_i p_i^2 )^2}{<}0.3$ 
which ensures reduced discretisation effects. This is based on empirical 
arguments~\cite{Constantinou:2010gr}, as this ratio appears to
suppress ${\cal O}(a^2)$ 
terms in the perturbative expressions for the Greens functions.
The procedure we follow in this work is the same as our previous work in 
non-perturbative renormalisation, and thus, we refer the interested reader to Refs.~\cite{Alexandrou:2010me,Alexandrou:2012mt,Alexandrou:2015sea} for technical 
details. It is worth mentioning that in the renormalisation of the one-derivative 
operator we also employ improvements by subtracting lattice artifacts~\cite{Alexandrou:2015sea}. 
The latter are computed to one-loop in perturbation theory and to all orders in the 
lattice spacing, ${\cal O}(g^2\,a^\infty)$. These artifacts are present in the non-perturbative 
vertex functions of the fermion propagator and fermion operators under study. Such an improvement 
is not yet available for the two-derivative operator, but finite $a$ effects are partly removed 
upon the $(a p)^2{\to}0$ extrapolation. In this Section we focus on the results for $Z_{\rm vDD}$, 
which are presented for the first time, while results on $Z_{\rm vD}$ have been extracted within 
the work of Ref.~\cite{Alexandrou:2015sea}.

\begin{figure}[th]
  \centering
  \includegraphics[width=.7\linewidth]{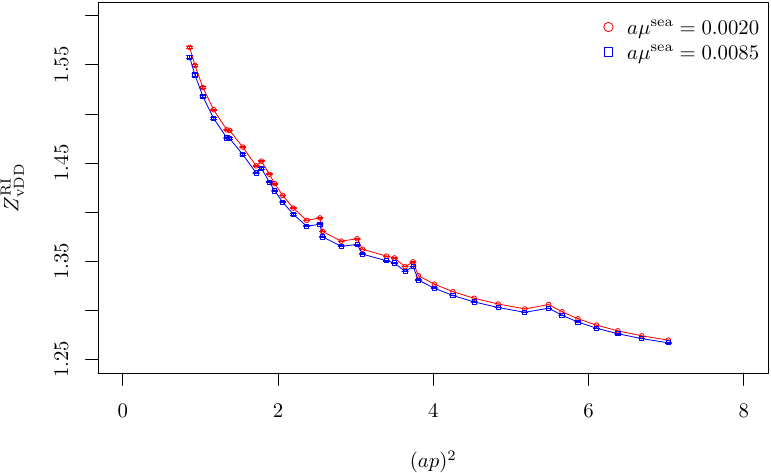}
  \caption{Pion mass dependence of $Z_{\rm vDD}^{{\rm RI}'}$ at $\beta{=}1.95$ as a function of the initial RI$'$ renormalisation scale ($p{=}\mu_0$).}
  \label{fig:ZvDDmpi}
\end{figure}

To extract the renormalisation functions in the chiral limit we perform an extrapolation using a 
quadratic fit with respect to the pion mass of the ensemble, that is, $a^{\rm{RI}'}\hspace{-0.1cm}(\mu_0)+b^{\rm{RI}'}\hspace{-0.1cm}(\mu_0) \hspace{-0.02cm} \cdot \hspace{-0.02cm} M_\pi^2$\, where 
$a$ and $b$ depend on the scheme and scale. In addition, these parameters depend on the coupling 
constant, and separate fits are performed at each value of $\beta$.
We find that the renormalisation functions under study have a very mild dependence on the pion mass, 
which becomes slightly larger for $(a\,p)^2{<}1$. However, these points do not participate in the fit 
$(a p)^2{\to}0$ for the final estimates. Allowing a slope, $b{\neq}0$, and performing a linear 
extrapolation with respect to $M_\pi^2$ the data yield a slope that is compatible with zero within the 
small uncertainties. This is demonstrated in Fig.~\ref{fig:ZvDDmpi}  for $Z^{\rm RI'}_{\rm vDD}$ for 
$\beta{=}1.95$, plotted as a function of the initial scale $(a p)^2$. For clarity we only show two 
values of the twisted mass $a\,\mu^{\rm sea}$, while the statistical errors are too small to be visible.
The corresponding plot for $Z_{\rm vD}$ is shown in Ref.~\cite{Alexandrou:2015sea}.

\begin{figure}[th]
  \centering 
  \includegraphics[width=0.7\linewidth]{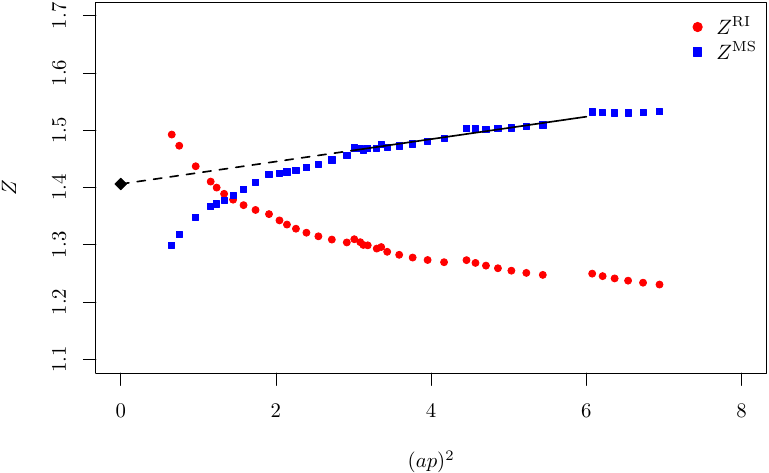}
  \caption{Chirally extrapolated renormalisation function for $\langle
    x^2 \rangle$ in the RI$'$ ($Z_{\rm vDD}^{{\rm RI}'}(\mu_0)$) and
    $\overline{\rm MS}$ ($Z_{\rm vDD}^{\overline{\rm MS}}(2\ {\rm GeV})$) 
    schemes at $\beta{=}2.10$, as a function of the initial renormalisation 
    scale ($p{=}\mu_0$). A black diamond represent the final estimate upon 
    $(a p)^2{\to}0$ and the solid line to fit in the interval $[3,6]$
    of $Z^\mathrm{MS}$.}
  \label{fig:ZRIMS}
\end{figure}

In order to compare lattice values to experimental results one must
convert to a universal renormalisation scheme and use a reference
scale $\overline{\mu}$. Typically one chooses the ${\overline{\rm
    MS}}$-scheme at $\overline{\mu}{=}2\ \mathrm{GeV}$. The conversion
from RI$'$ to ${\overline{\rm MS}}$ scheme uses the intermediate
Renormalization Group Invariant (RGI) scheme, which is scale
independent and thus, 
\begin{equation}
Z^{\rm RGI}_{\cal O} =
Z_{\cal O}^{\mbox{\scriptsize RI$^{\prime}$}} (\mu_0) \, 
\Delta Z_{\cal O}^{\mbox{\scriptsize RI$^{\prime}$}}(\mu_0) =
Z_{\cal O}^{\overline{\rm MS}} (2\,{\rm GeV}) \,
\Delta Z_{\cal O}^{\overline{\rm MS}} (2\,{\rm GeV})\,.
\end{equation}
The conversion factor can be extracted from the above relation
\begin{equation}
C_{\cal O}^{{\rm RI}',{\overline{\rm MS}}}(\mu_0,2\,{\rm GeV}) \equiv 
\frac{Z_{\cal O}^{\overline{\rm MS}} (2\,{\rm GeV})}{Z_{\cal O}^{{\rm RI}'} (\mu_0)} = 
\frac{\Delta Z_{\cal O}^{\mbox{\scriptsize RI$^{\prime}$}}(\mu_0)}
     {\Delta Z_{\cal O}^{\overline{\rm MS}}(2\,{\rm  GeV})}\,.
\end{equation}
The quantity $\Delta Z_{\cal O}^{\mathcal S}(\mu_0)$ is expressed in terms of the 
$\beta$-function and the anomalous dimension $\gamma_{\cal O}^S \equiv \gamma^S$ 
of the operator
\begin{equation}
\Delta Z_{\cal O}^{\mathcal S} (\mu) =
  \left( 2 \beta_0 \frac {{g^{\mathcal S} (\mu)}^2}{16 \pi^2}\right)
^{-\frac{\gamma_0}{2 \beta_0}}
 \exp \left \{ \int_0^{g^{\mathcal S} (\mu)} \! \mathrm d g'
  \left( \frac{\gamma^{\mathcal S}(g')}{\beta^{\mathcal S} (g')}
   + \frac{\gamma_0}{\beta_0 \, g'} \right) \right \}\,,
\end{equation}
with all necessary ingredients defined in Appendix~\ref{appA}. We employ a 3-loop approximation, 
for which $\Delta Z_{\cal O}^{\mathcal S} (\mu_0)$ takes a simpler form~\cite{Alexandrou:2015sea}.

\begin{table}[th]
  \begin{center}
    \begin{tabular}{clll}
      \hline
      \hline
      RFs &  $\quad \beta{=}2.10$ & $\quad  \beta{=}1.95$ & $\quad \beta{=}1.90$  \\[0.21ex]
      \hline
      \\[-2.5ex]
      $Z_{\rm vD}^{\overline{\rm MS}}$      &$\quad$1.0991(29)(55)$\quad$          &$\quad$1.0624(108)(33)$\quad$            &$\quad$1.0268(26)(103)$\quad$ \\[0.21ex]
      $Z_{\rm vDD}^{\overline{\rm MS}}$   &$\quad$1.406(1)(20)$\quad$                          &$\quad$1.356(1)(18)$\quad$           &$\quad$1.307(1)(21)$\quad$     \\[0.2ex]
      \hline
      \hline
    \end{tabular}
    \caption{Our final values of the renormalisation functions $Z_{\rm
        vD}^{\overline{\rm MS}}$ and  $Z_{\rm vDD}^{\overline{\rm
          MS}}$ at $\bar\mu=2$~GeV renormalisation scale. The first error is the
      statistical error. The second error corresponds to
      the systematic error obtained by varying the 
      fit range in the $(a\,p)^2{\to} 0$ extrapolation.} 
    \label{tabZ}
  \end{center}
\end{table}

In Fig.~\ref{fig:ZRIMS} we present representative results for $Z_{\rm
  vDD}$ (at $\beta{=}2.10$) in the RI$'$ ($Z^{\rm RI'}_{\rm
  vDD}(\mu_0)$) and ${\overline{\rm MS}}$ ($Z^{\overline{\rm MS}}_{\rm
  vDD}(2\ {\rm GeV})$) schemes as a function of the initial RI renormalisation 
  scale, $\mu_0{=}p$. Note that $Z^{\overline{\rm MS}}_{\rm vDD}$ has been 
  evolved to 2~GeV, but there is residual dependence on the initial scale. 
  This dependence is removed by extrapolating to zero, using the Ansatz
\begin{equation}
Z_{\cal O}(a\,p) = Z_{\cal O}^{(0)} + Z_{\cal O}^{(1)}\cdot(a\,p)^2\,,
\label{Zfinal}
\end{equation}
where $ Z_{\cal O}^{(0)}$ corresponds to our final value on the renormalisation 
functions for the operator ${\cal O}$. For each value of $\beta$ we consider 
momenta $6\, \ge\, (a\,p)^2\, {\ge}\, 2$ for which perturbation theory is trustworthy and 
lattice artifacts are still under control.

In Table~\ref{tabZ} we report our chirally extrapolated values for
$Z_{\rm vD}^{(0)}$ and $Z_{\rm vvD}^{(0)}$ in the ${\overline{\rm
    MS}}$ scheme at 2~GeV. $Z_{\rm vD}^{(0)}$ has been extracted upon
subtraction of the ${\cal O}(g^2\,a^\infty)$ terms, which improves the
estimates as explained in Ref.~\cite{Alexandrou:2015sea}. The
statistical and systematic uncertainties are given in the first and
second parentheses, respectively. For $Z_{\rm vvD}^{(0)}$ we chose as
appropriate fit interval $(a p)^2: [3-6]$. The reported systematic 
uncertainty is extracted from the difference of $Z_{\rm vvD}^{(0)}$ 
between various intervals for the $(a\,p)^2{\to} 0$ extrapolation.

In Ref.~\cite{Blossier:2014kta} $Z_{\rm vD}$ has been determined on
the same gauge configurations using a different method. The authors
find generally lower values for $Z_{\rm vD}$, which are within the
quoted systematic uncertainties compatible with what we quote in
Table~\ref{tabZ}. In order to be consistent in our treatment of
$\langle x\rangle$ and $\langle x^2\rangle$, we stick here to the
values compiled in Table~\ref{tabZ}.

%% file: results.tex
\section{Results}
\label{sec:results}

\begin{figure}[t]
  \centering
  \subfigure{\includegraphics[width=.48\linewidth,page=3]
    {plots/A100sample}}\quad
  \subfigure{\includegraphics[width=.48\linewidth,page=1]
    {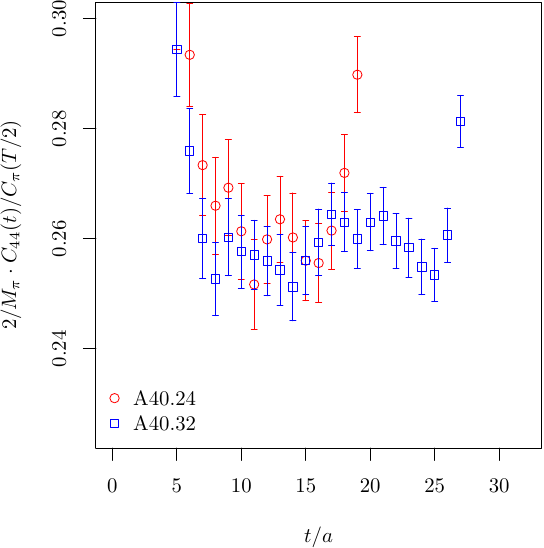}}\quad
  \caption{Left: The bare three point function $C_{44}(t)$ for ensemble
    D30.48 as a function of $t/a$. Right: The bare $\langle
    x\rangle(t)$ as a function of $t/a$ for A40.24 and A40.32.}
  \label{fig:fs}
\end{figure}

In the left panel of Figure~\ref{fig:fs} we show the bare three-point
function $C_{44}(t)$ for ensemble D30.48. The plot demonstrates the
quality of the data we are able to obtain for $C_{44}(t)$. 
In Ref.~\cite{Guagnelli:2004ww} it was found that, in the quenched
approximation, with Schr{\"o}dinger functional boundary conditions and
clover improved Wilson fermions,
 finite size effects to $\langle x\rangle$ are quite 
sizable. This persists even at values of $M_\pi\cdot L$ where finite 
size effects in $M_\pi$ are no longer visible. The authors measured 
these effects to be of about 5\% at values of $M_\pi\cdot L\approx4$.
In this work we use two ensembles, A40.24 with $M_\pi\cdot L=3.5$ and
A40.32 with $M_\pi\cdot L=4.5$, which differ 
only in the volume, and can be used for investigation of finite size 
effects. In the right panel of Figure~\ref{fig:fs} we show a comparison 
of $\langle x\rangle(t)$ between A40.24 and A40.32. We find that the
values of the bare $\langle x\rangle(t)$ in the plateau regions for A40.24 and 
A40.32 agree within error bars. This indicates that in our lattice 
discretisation for the given values of $M_\pi\cdot L$, finite size
effects play a minor role, if any. This is in agreement with the
finding in the quenched approximation~\cite{Capitani:2005aa}.

\begin{figure}[t]
  \centering
  \subfigure{\includegraphics[width=.48\linewidth,page=1]
    {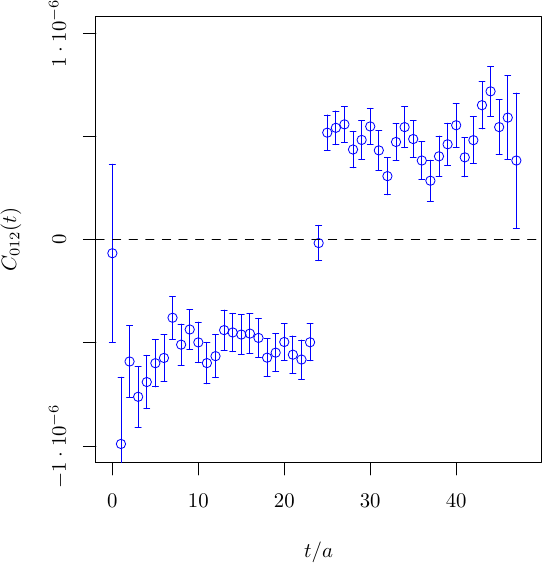}}\quad
  \subfigure{\includegraphics[width=.48\linewidth,page=2]
    {plots/A100sample}}\quad
  \caption{The bare three point function $C_{012}(t)$ as a function of
    $t/a$ for ensemble A100.24 (left) and for ensemble A30.32 (right).}
  \label{fig:bare012}
\end{figure}

In Figure~\ref{fig:bare012} we show two examples for the bare data of
the three-point function $C_{012}$ for the ensembles A100.24 in the
left panel and A30.32 in the right panel. One nicely observes the
asymmetry in $C_{012}$ around $t=T/2$. The signal to noise ratio
deteriorates significantly with decreasing light quark mass value. 
Compared to $C_{44}$ a strong increase in the statistical uncertainty
is clearly visible. Still, the determination of $\langle x^2\rangle$
is feasible for all quark mass values. Finite size effects for the
case of $\langle x^2\rangle$ are within the reported statistical 
uncertainties.

For determining an estimate of $\langle x\rangle$ and $\langle
x^2\rangle$ we perform plateau fits to the (anti-) symmetrized data
for $\langle x\rangle(t)$ and $\langle x^2\rangle(t)$. Following the
ideas put forward in Ref.~\cite{Helmes:2015gla}, we perform such
fits for many different fit ranges. The estimates for $M_\pi$ and
$f_\pi$ are obtained by fitting all possible fit ranges with at least
6 consecutive time slices. Each of these fits obtains a weight
according to
\begin{equation}
  \label{eq:fitweight}
  w_i\ =\ \frac{(1 - 2|p_i - 1/2|)^2}{\Delta_i^2}\,.
\end{equation}
Here, $p_i$ is the $p$-value of the corresponding fit and $\Delta_i$
is the statistical error on $M_\pi$ or $f_\pi$ determined from the
bootstrap procedure for this fit range. This procedure is repeated for
each bootstrap replica. In addition, a systematic uncertainty from
the fit range choice can be
specified from the $68\%$ confidence interval of the weighted distribution.

The estimates for the moments are then obtained in a very similar
manner, just that also $M_\pi$ is needed. Thus, we combine all possible fit
ranges with at least six consecutive time slices for $C_\pi$ with all
possible fit ranges with at least six consecutive time slices to the
corresponding three point function. The weight for a moment with a
specific fit range combination is obtained by multiplying the
corresponding weights of the fit to $C_\pi$ and the fit to the 
three point function.

The estimates extracted as explained above for the first and second moment are
compiled in Table~\ref{tab:results211}. The values are renormalised at
$2\ \mathrm{GeV}$ in the $\overline{\mathrm{MS}}$-scheme. Statistical
errors coming from the renormalisation functions are included via the
parametric bootstrap procedure. The second error quoted comes from the
different fit ranges estimated as discussed before. One observes that
this systematic uncertainty is for the first moment usually of the
order of the statistical error. For the second moment it is sometimes a
bit larger. D15.48 and B25.32 have, unfortunately, a large statistical
and systematic uncertainty on $\langle x^2\rangle$. In particular for
D15.48 the significance of the result strongly depends on the chosen
fit range. The reason is the significant increase of noise towards
smaller light quark mass values. 

\begin{table}[t!]
  \centering
  \begin{tabular*}{.7\textwidth}{@{\extracolsep{\fill}}lcc}
    \hline\hline
    ensemble & $\langle x\rangle_\mathrm{R}$
    & $\langle x^2\rangle_\mathrm{R}$\\
    \hline\hline
    A30.32  & \(0.2586(41)(28)\) & \(0.131(18)(24)\)\\  
    A40.24  & \(0.2630(44)(16)\) & \(0.116(20)(26)\)\\  
    A40.32  & \(0.2652(37)(26)\) & \(0.114(16)(29)\)\\  
    A60.24  & \(0.2782(36)(17)\) & \(0.116(15)(08)\)\\ 
    A80.24  & \(0.2835(33)(10)\) & \(0.115(10)(08)\)\\ 
    A100.24 & \(0.2921(33)(05)\) & \(0.123(08)(08)\)\\
    \hline                                                                   
    B25.32  & \(0.2523(51)(71)\) & \(0.132(40)(53)\)\\  
    B35.32  & \(0.2617(41)(33)\) & \(0.109(21)(28)\)\\  
    B55.32  & \(0.2770(36)(17)\) & \(0.134(12)(16)\)\\  
    B85.24  & \(0.2902(35)(47)\) & \(0.139(09)(07)\)\\
    \hline                                                                   
    D15.48  & \(0.2331(50)(32)\) & \(0.18(06)(20)\)\\   
    D30.48  & \(0.2510(25)(37)\) & \(0.122(20)(38)\)\\  
    D45.32  & \(0.2610(31)(20)\) & \(0.153(14)(12)\)\\  
    \hline\hline
    \vspace*{0.1cm}
  \end{tabular*}
  \caption{The results for the renormalised $\langle
    x\rangle_\mathrm{R}$ and $\langle x^2\rangle_\mathrm{R}$ for the
    ensembles used in this investigation. $\langle
    x\rangle_\mathrm{R}$ and $\langle x^2\rangle_\mathrm{R}$ are given
    at $\bar\mu = 2\ \mathrm{GeV}$ in the $\overline{\mathrm{MS}}$-scheme.}
  \label{tab:results211}
\end{table}

\begin{figure}[t]
  \centering
  \subfigure{\includegraphics[width=.48\linewidth,page=1]
    {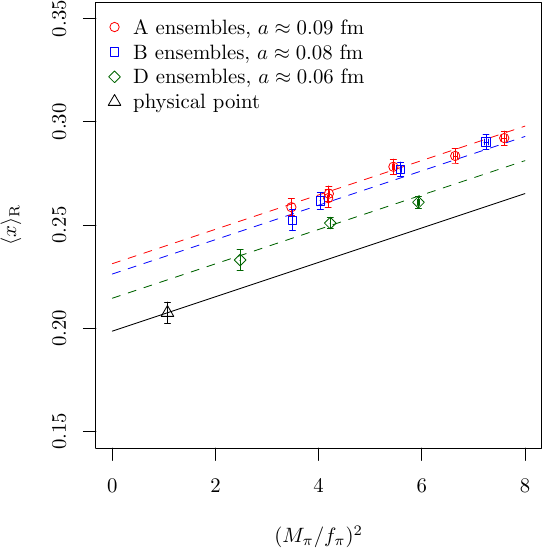}}\quad
  \subfigure{\includegraphics[width=.48\linewidth,page=1]
    {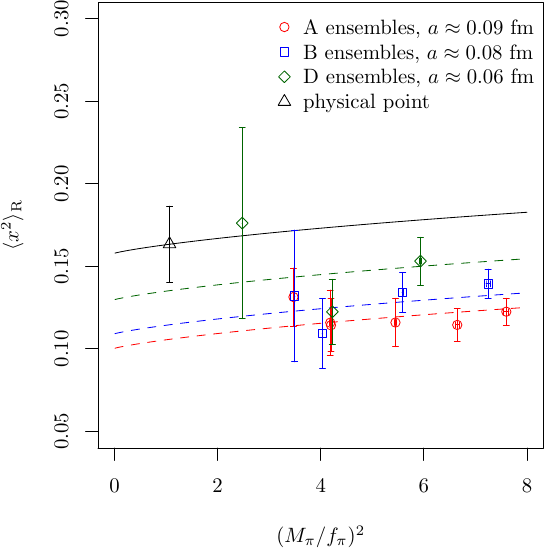}}\quad

  \caption{$\langle x\rangle_\mathrm{R}$ of the pion (left)
    and $\langle x^2\rangle_\mathrm{R}$ (right) as functions of $(M_\pi/f_\pi)^2$ at
    $\bar\mu=2\ \mathrm{GeV}$ in the 
    $\overline{\mathrm{MS}}$ scheme.
    Dashed colored lines represent the best fit
    functions Eq.~\ref{eq:fits} at the three lattice spacing values, respectively. The
    solid black line represents the continuum curve. The black
    triangles represent the estimates at the physical point in the
    continuum limit. The error bars represent only the statistical
    uncertainty.}
  \label{fig:averx211}
\end{figure}

These results for the renormalised first and second moments of the pion
are shown in Figure~\ref{fig:averx211} in the left and right panel,
respectively. They are plotted as a function of $(M_\pi/f_\pi)^2$ with
statistical errors only.

\subsection{Chiral and Continuum Extrapolations}

The ChPT expressions Eq.~\ref{eq:ChPTx} and \ref{eq:ChPTx2} plus
terms proportional to $(a/r_0)^2$ for $\langle x\rangle$ and $\langle
x^2\rangle$ read
\begin{equation}
  \label{eq:fits}
  \begin{split}
    \langle x\rangle_\mathrm{bare} [(M_\pi/f_\pi)^2; \beta]\ &=\ \frac{1}{Z_\mathrm{vD}(\beta)}\left[c_0 + c_1 \frac{M_\pi^2}{f_\pi^2}\right] +
    c_a(a/r_0(\beta))^2\,,\\
    \langle x^2\rangle_\mathrm{bare} [(M_\pi/f_\pi)^2; \beta]\ &=\ \frac{1}{Z_\mathrm{vDD}(\beta)}\left[b_0\left(1 - \frac{M_\pi^2}{(4\pi
f_\pi)^2}\log\frac{M_\pi^2}{f_\pi^2}\right) + b_1
    \frac{M_\pi^2}{f_\pi^2}\right] + b_a(a/r_0(\beta))^2\,.\\
  \end{split}
\end{equation}
We perform fits of these functional forms to all the data
of the first and second moment separately. For these fits we have
the data for 
the bare $\langle x\rangle$ ($\langle x^2\rangle$) and the estimates
for $Z_\mathrm{vD}$ ($Z_\mathrm{vDD}$) and $r_0/a$. To properly account for the
uncertainties in the renormalisation functions and the Sommer
parameter $r_0/a$, we use the augmented $\chi^2$ function as follows
\begin{equation}
  \chi^2_\mathrm{aug}\ =\ \chi^2 +
  \sum_\beta\left(\frac{Z(\beta)-P_Z(\beta)}{\Delta Z(\beta)}\right)^2 +
  \sum_\beta\left(\frac{\frac{r_0}{a}(\beta)-P_{r_0}(\beta)}{\Delta \frac{r_0}{a}(\beta)}\right)^2\,.
\end{equation}
Here, $Z$ and $\Delta Z$ denote the relevant renormalisation factor
and its statistical uncertainty either for the first or the second
moment. $P_Z$ and $P_{r_0}$ are additional fit parameters per
$\beta$-value. The usual $\chi^2$ function entering
$\chi^2_\mathrm{aug}$ reads
\begin{equation}
  \chi^2\ =\ \sum_\beta\ \sum_{i(\beta)}
  \left(\frac{y_i - g(x_i, \{P\})}{\Delta y_i} \right)^2\,.
\end{equation}
Here $i(\beta)$ index the data points for the corresponding
$\beta$-value, $y_i$ is the bare data for $\langle x\rangle$ ($\langle
x^2\rangle$) and $x_i$ the data for $(M_\pi/f_\pi)^2$. With $\{P\}$ we
label the set of fit parameters $\{c_0, c_1, c_a, P_{r_0}, P_{Z_\mathrm{vD}}\}$ ($\{b_0, b_1,
b_a, P_{r_0}, P_{Z_\mathrm{vDD}}\}$) and with $g$ the corresponding ChPT expression.
The equation for the $\chi^2$ function above is written for the
uncorrelated case, because all data points stem from independent
ensembles and $r_0/a$ and the renormalisation constants from
independent analyses. Errors of fit parameters
are again computed using the bootstrap procedure by performing a fit
on every bootstrap replica.

In principle one could also include the error on $(M_\pi/f_\pi)^2$ in
the fit. However, these errors are so small compared to the ones for
the moments that they do not alter the fit results. We also do not
include systematic uncertainties in the fit, because they lack a
statistical interpretation and would increase all error
bars more or less uniformly.

For the first moment we obtain the following best fit parameters
\begin{equation}
  c_0 = 0.199(5)\,,\quad c_1 = 0.0083(5)\,,\quad c_a = 0.92(20)\,.
\end{equation}
The $p$-value of the fit equals $0.61$ with $\chi^2_\mathrm{aug}/\mathrm{dof} =
8.2/10$. Thus, the fit is acceptable and the continuum value of
$\langle x\rangle$ at the physical point -- defined via
$M_\pi/f_\pi=1.0337$ -- reads 
\begin{equation}
  \label{eq:resx}
  \langle x\rangle_\mathrm{R}^\mathrm{phys}\ =\ 0.2075(53)\,.
\end{equation}
The best fit curves for the three lattice spacings are included as
dashed lines in the left panel of Figure~\ref{fig:averx211}. The
continuum curve is the black solid line with the estimate of the
first moment at the physical point indicated with the black triangle.

If we had used the values for $Z_\mathrm{vD}$ from
Ref.~\cite{Blossier:2014kta} instead of the ones we quote in
Table~\ref{tabZ}, we had obtained an equally good fit with result
$\langle x\rangle_\mathrm{R}^\mathrm{phys}=0.189(16)$. This value is
compatible with the value above, but with larger uncertainties. We
include half of the difference as systematic error in our final value.

For the second moment, the best fit parameters read
\begin{equation}
  b_0 = 0.16(2)\,,\quad b_1 = 0.005(2)\,,\quad b_a = -1.6(7)\,
\end{equation}
With a $p$-value of $0.89$ ($\chi^2_\mathrm{aug}/\mathrm{dof}=5/10$)
the continuum estimate at the physical 
point reads
\begin{equation}
  \label{eq:resx2}
  \langle x^2\rangle_\mathrm{R}^\mathrm{phys}\ =\ 0.163(23)
\end{equation}
Like for $\langle x\rangle_\mathrm{R}$, the corresponding curves are
shown in 
the right panel of Figure~\ref{fig:averx211} in addition to the
data. Again, the black triangle represents the estimate of the second
moment at the physical point in the continuum limit.

For both the first and the second moment the fit parameters for the
renormalisation factors and for $r_0/a$ agree very well with the
input data. All best fit parameters, their uncertainties and
correlations are compiled in Appendix~\ref{sec:appB}.

As is visible from Figure~\ref{fig:averx211} and from the $p$-values,
the data is well described by the ChPT expressions in the full range
of pion mass values we have available. However, it is questionable
whether one-loop ChPT works for pion masses up to about
$500\ \mathrm{MeV}$. Therefore, we have repeated the fits excluding
all data points with $(M_\pi/f_\pi)^2 > 6$. The so obtained results
are well compatible within error bars with the results quoted
above. Also the $p$-values of the fits do not improve. Thus, we
conclude that our statistical uncertainty covers this
systematics. This point needs to be reconsidered once NLO formulae
are available.

%% file: discussion.tex
\section{Discussion}
\label{sec:discussion}

In this work we demonstrate the feasibility of the lattice calculation 
for the first and second moments of the pion PDF. Despite the challenges
present in calculations of higher moments, we find sufficiently long 
plateau regions for the bare matrix elements for all ensembles used here,
with the dependence on the fit-range of the order of the statistical uncertainties.
For the first moment, where our bare values are precise to the
few percent level, we observe a sizable dependence on $M_\pi^2$ and
significant lattice artifacts, cf. the left panel of
Figure~\ref{fig:averx211}. From the value of $\langle
x\rangle_\mathrm{R}$ of ensemble D15.48, which is the smallest pion
mass value closest to the continuum limit, there is still a 10\%
difference to the continuum value at the physical point.

The statistical errors for $\langle x^2\rangle$ are significantly larger than for
$\langle x\rangle$, since two derivatives and two non-zero spatial
components of momentum are required. Therefore, pion mass and lattice
spacing dependence are both not significant: all the data could be fitted 
to a constant in $M_\pi/f_\pi$ with a result similar to the one we quote
above. For both moments, finite size effects turn out to be not
relevant, which is in agreement with the finding of
Ref.~\cite{Abdel-Rehim:2015owa}, where the twisted mass formulation
was used as well.

Our results for the first moment can be compared to other lattice
computations, including our recent work using $N_f=2$ simulations 
directly at the physical point, however, without extrapolation to 
the continuum limit~\cite{Abdel-Rehim:2015owa}. The value found in
Ref.~\cite{Abdel-Rehim:2015owa} also neglecting disconnected
contributions at the physical point reads
$\langle x\rangle_\mathrm{R} = 0.214(15)(^{+12}_{-9})$. It is fully compatible with
the result we find here. In Refs.~\cite{Brommel:2006zz,Brommel:2007zz}
a value of $\langle x\rangle_\mathrm{R} = 0.271(2)(10)$ at $\bar\mu =
2\ \mathrm{GeV}$ in the $\overline{\mathrm{MS}}$ scheme is quoted for 
$N_f=2$ flavor QCD also neglecting disconnected diagrams, which is
significantly larger than our value. In these two references almost no
lattice artifacts appear to be visible, in contrast to our findings.
In the work of Bali et al.~\cite{Bali:2013gya} a significantly lower value
is reported, using a single ensemble at near physical pion mass value.

It is not so easy to identify a reason for the differences we
observe. It seems the number of flavors is not so important, because
our result with $N_f=2+1+1$ quark flavors is fully compatible with
the $N_f=2$ result at the physical point. Even though the latter
computation is at a single lattice spacing only, lattice spacing
effects seem to be small with this
action~\cite{Abdel-Rehim:2015pwa}. Thus, differences are 
likely to come from the chosen lattice discretisation leading to different
lattice artifacts and finite size effects. This clearly demands furher
careful investigations of systematic uncertainties in the future.

Refs.~\cite{Brommel:2006zz,Brommel:2007zz} present the calculation for
$\langle x^2\rangle$ using a different operator that possibly mixes
under renormalisation. The authors compute only the connected diagram,
too, and find $\langle x^2\rangle_\mathrm{R} =
0.128(9)(4)$ at $\bar\mu = 2\ \mathrm{GeV}$ in the $\overline{\mathrm{MS}}$
scheme, compatible with our result.

It is utterly important to relate the values of the moments computed
in this paper to what is measured experimentally. But, in our
computation fermionic disconnected contributions to the three-point
functions $C_{44}$ and $C_{012}$ have been neglected. Thus, strictly
speaking from a quantum field theory point of view, the spectral
decomposition of the (connected only) three-point functions is not
possible. A meaning is only recovered if we rely on the assumption
that the fermionic disconnected contributions have a negligible share
to the total three-point functions.

On the other hand, in practice the fermionic connected and
eventually also the disconnected contributions can be determined. 
It is then very appealing to identify the part coming from the
disconnected 
contributions as purely sea moments, see fig.~\ref{fig:diagrams}. This
allows one to make contact to the phenomenological point of view, where
typically the following sum rule is used, here for $\langle x\rangle$ 
\begin{equation}
  \label{sum_rule_pheno}
  2 \langle x \rangle_v (\bar\mu^2) + 2 N_f \langle x \rangle_s
  (\bar\mu^2) + \langle x \rangle_G (\bar\mu^2)  = 1, 
\end{equation}
where the $v, s, G$ denote the valence quark, sea quark, and gluon
contributions, respectively. On the other hand, for a lattice
calculation one would write:
\begin{equation}
\label{sum_rule_lattice}
\langle x \rangle_u^\textrm{conn} (\bar\mu^2) + \langle x
\rangle_d^\textrm{conn} (\bar\mu^2) + \sum_q \langle x
\rangle_q^\textrm{disc} (\bar\mu^2) +  \langle x \rangle_G (\bar\mu^2)
= 1, 
\end{equation}
where $\textrm{conn}$ ($\textrm{disc}$) stands for a lattice
computation performed with only fermionic connected (disconnected)
contributions to the corresponding three-point function taken into
account. The sum in $q$ is over all active quark flavors. As defined
in Eq.~\ref{eq:O44}, the quantity calculated in this work is the
total connected only contribution: 
\begin{equation}
  \label{renor_average_x}
  \langle x \rangle_\textrm{R}^\textrm{phys}(\bar\mu^2)\ =\ \langle x
  \rangle_u^\textrm{conn} (\bar\mu^2)\ =\ \langle x
  \rangle_d^\textrm{conn} (\bar\mu^2). 
\end{equation}
Still, from Eq.~\ref{renor_average_x} it is clear that $\langle x
\rangle_\textrm{R}^\textrm{phys}(\bar\mu^2)$ cannot be the valence
contribution of Eq.~\ref{sum_rule_pheno}, because the connected
contributions also receive contributions from so-called Z diagrams,
which are counted as sea quark distributions in Eq.~\ref{sum_rule_pheno}. 
Nevertheless, since the following equality must hold: 
\begin{equation}
  \langle x \rangle_u^\textrm{conn} (\bar\mu^2) + \langle x
  \rangle_d^\textrm{conn} (\bar\mu^2) + \sum_q \langle x
  \rangle_q^\textrm{disc} (\bar\mu^2) = 
2 \langle x \rangle_v (\bar\mu^2) + 2 N_f \langle x \rangle_s (\bar\mu^2),
\end{equation}
we may, keeping the caveat discussed above in mind,
compare $\langle x \rangle_\mathrm{R}^\mathrm{phys}(\bar\mu^2)$ with
phenomenology if we understand the quantity computed here as an upper
limit for $\langle x \rangle_v (\bar\mu^2)$.

Phenomenological results for average $x$ and $x^2$
are provided in Refs.~\cite{Barry:2018ort} and~\cite{Wijesooriya:2005ir}. 
Below we compare to the more recent results from Ref.~\cite{Barry:2018ort}, 
which are based on a larger set of experimental data, where they find
\begin{equation}
  2\langle x\rangle_v\ =\ 0.49(1)\,,\qquad2\langle x^2\rangle_v\ =\ 0.217(4)\,,
\end{equation}
both in the $\overline{\mathrm{MS}}$ scheme at
$\bar\mu=2\ \mathrm{GeV}$. Compared to our results in
Eq.~\ref{eq:resx} and Eq.~\ref{eq:resx2}, i.e. $2\langle x
\rangle_u^\textrm{conn}$ and $2\langle x^2 \rangle_u^\textrm{conn}$,
respectively, we observe a tension for $\langle x\rangle$.
In particular, the value for $\langle x \rangle$ we observe
is smaller than the phenomenological estimate, which is opposite to
what we expect from our discussion above. This tension might be
explained with the caveats lined out above, noticing also that
according to Ref.~\cite{Barry:2018ort}, the extraction of $\langle x
\rangle_v$ is still sensitive to the inclusion of new data sets, being
reduced when leading neutron production data is added to previously
existing Drell-Yan data. The results we find here point to the
direction of further reductions of $\langle x \rangle_v$. in this
context, experimental efforts planned at
COMPASS~\cite{COMPASS,FrancoCOMPASS} and JLab~\cite{JLAB} to measure
the pion structure functions will be instrumental to settle this
matter, having also an impact in the decomposition of the pion
momentum sum rule. Our value for $\langle x^2\rangle$ is larger than
$2\langle x^2\rangle_v$, but its also has a large error bar.

Finally, we note that the relative share of connected to disconnected
contribution to the total $\langle x\rangle$ may well depend on the
pion mass.

%% file: summary.tex
\section{Summary}
\label{sec:summary}

In this paper we have presented results for the first and second
moment of the pion PDF computed in $N_f=2+1+1$ lattice QCD. While we
still neglect fermionic disconnected diagrams for both moments, we
have thoroughly investigated the extrapolations to the physical point and to
the continuum. This was possible due to ETMC ensembles spanning three
values of the lattice spacing and pion masses ranging from 270 to 500 MeV.
For $\langle x\rangle$ and $\langle x^2\rangle$ we use operators which
avoid any mixing under renormalisation. By studying two ensembles with
all identical parameters but the lattice size, we can exclude finite
volume effects significantly larger than our statistical
uncertainties.

For the computation of $\langle x^2\rangle$ non-zero spatial momenta
are required which we inject using twisted boundary conditions. These
allow us to chose the momenta optimally for the signal to noise ratio in
the corresponding three-point function. Still, our results for
$\langle x^2\rangle$ have significantly larger statistical
uncertainties than the ones for $\langle x\rangle$, which is of course
also due to the second derivative needed for $\langle x^2\rangle$. 

It turns out that the choice of fit-ranges represents a major
systematic uncertainty in the calculation of the moments. We approach
this uncertainty by performing many fits and include them all weighted
appropriately in the final estimates. From the weighted distribution a systematic
error can be estimated which is typically of the order of the
statistical error. The only exception is our ensemble at the smallest
lattice spacing and pion mass value, where the systematic errors
prevent us from obtaining a significant result.
In summary we obtain
\[
\langle x\rangle_\mathrm{R}^\mathrm{phys}=0.2075(53)_\mathrm{stat}(20)_\mathrm{sys}(90)_\mathrm{Z}\quad
\textrm{and}\quad \langle 
x^2\rangle_\mathrm{R}^\mathrm{phys} = 0.163(23)_\mathrm{stat}(25)_\mathrm{sys}\,,
\]
determined at $2\ \mathrm{GeV}$ in the
$\overline{\mathrm{MS}}$-scheme. In the bare matrix elements we find
on average a 1\% systematic error on $\langle x\rangle$ and a 15\%
systematic error on $\langle x^2\rangle$, which we have added to the
final results in order to reflect the
systematic uncertainty coming from the fit range choice. In $\langle
x\rangle$ we add the systematic uncertainty from using the $Z$-factors
determined in Ref.~\cite{Blossier:2014kta} instead of the ones
compiled in Table~\ref{tabZ}.

The comparison to phenomenology is difficult, because in our
computation fermionic disconnected contributions to the three-point
functions have been neglected. However, if one identifies the
quantities computed here with an upper limit to what is called valence
contribution in phenomenology, we observe that our value for $\langle
x\rangle$ is smaller compared to phenomenology, while the value for
$\langle x^2\rangle$ is also larger compared to phenomenology, but has
large error bars.

From the discussion in the previous section it is clear that a
computation including fermionic disconnected diagrams is highly desirable.
Thus, we are planning to repeat this computation by including
fermionic disconnected contributions to the three point functions. Then
also the gluonic moments ought to be computed to properly perform the
renormalisation procedure.

\section*{Acknowledgments}

We thank the
members of ETMC for the most enjoyable collaboration. We thank
G.~Bali, M.~Mangin Brinet and I.~Schienbein for valueable
discussions. CU thanks the LPSC Grenoble, where
part of this work has been carried out, for the kind hospitality.
MC acknowledges financial support by the U.S. National Science 
Foundation under Grant No.\ PHY-1714407.
The computer time for this project was made available to us by the
John von Neumann-Institute for Computing (NIC) on the Juqueen and
Jureca systems in J{\"u}lich. 
This project was funded by the DFG as a project with number 392578569
and in parts in the Sino-German CRC110. The open source software
packages tmLQCD~\cite{Jansen:2009xp}, Lemon~\cite{Deuzeman:2011wz},
QUDA~\cite{Clark:2009wm,Babich:2011np,Clark:2016rdz} and 
R~\cite{R:2005} have been used. SZ acknowledges support by the DFG Collaborative
Research Centre SFB 1225 (ISOQUANT).

%% file: appendix.tex
\section{$\beta-$function and anomalous dimensions}
\label{appA}

In this Appendix we provide the definition of the $\beta-$function and the anomalous dimension of the two operators
presented in this work. To simplify the expressions we give the perturbative coefficients in the Landau gauge and in $SU(3)$.

The perturbative expansion of the anomalous dimension in a renormalisation scheme $\mathcal S$ is given as follows:
\begin{equation}
  \gamma^{\mathcal S} = - \mu \frac{\mathrm d}{\mathrm d \mu}
  \log Z_{\mathcal S} =
  \gamma_0 \frac{g^{\mathcal S} (\mu)^2}{16 \pi^2}
  + \gamma_1^{\mathcal S}
  \left( \frac{g^{\mathcal S} (\mu)^2}{16 \pi^2} \right)^2
  + \gamma_2^{\mathcal S}
  \left( \frac{g^{\mathcal S} (\mu)^2}{16 \pi^2} \right)^3
  + \cdots\,,
  \label{gammaS}
\end{equation}
while the $\beta-$function is defined as:
\begin{equation}
  \beta^{\mathcal S} =  \mu \frac{\mathrm d}{\mathrm d \mu}
  g^{\mathcal S} (\mu) =
  - \beta_0 \frac{g^{\mathcal S} (\mu)^3}{16 \pi^2}
  - \beta_1 \frac{g^{\mathcal S} (\mu)^5}{(16 \pi^2)^2}
  - \beta_2^{\mathcal S} \frac{g^{\mathcal S} (\mu)^7}{(16 \pi^2)^3}
  + \cdots\,.
\end{equation}
For the conversion from the RI$'$ to the ${\overline{\rm MS}}$ scheme
we use the three-loop expressions, to which the coefficients of the
$\beta-$function coincide and are given by~\cite{vanRitbergen:1997va,Gracey:2003yr}:
\begin{eqnarray}
  \beta_0 & = & 11 - \frac{2}{3} N_f \,, \\
  \beta_1 & = & 102 - \frac{38}{3} N_f \,, \\
  \beta_2 & = & \frac{2857}{2} - \frac{5033}{18} N_f
  + \frac{325}{54} N_f^2 \,.
\end{eqnarray}
All necessary expressions to convert to the ${\overline{\rm MS}}$ scheme are presented below. An upper index appears for scheme-dependent quantities, in order to denote the scheme that they correspond to.

{\underline{One-derivative vector/axial}}~\cite{Gracey:2003mr,Gockeler:2010yr}:
\begin{eqnarray}
  \gamma_0 & = & \frac{64}{9} \,, \\
  \gamma_1^{\overline{\rm MS}} & = & \frac{23488}{243} - \frac{512}{81} N_f \,, \\
  \gamma_1^{\mbox{\scriptsize RI$^{\prime}$}} & = & \frac{48040}{243} - \frac{112}{9} N_f \,, \\
  \gamma_2^{\overline{\rm MS}} & = & \frac{11028416}{6561} + \frac{2560}{81} \zeta_3
  - \left( \frac{334400}{2187}
  + \frac{2560}{27} \zeta_3 \right) N_f - \frac{1792}{729} N_f^2\,,\\
  \gamma_2^{\mbox{\scriptsize RI$^{\prime}$}} & = & \frac{59056304}{6561} 
  - \frac{103568}{81} \zeta_3 - \left(\frac{2491456}{2187}
  + \frac{416}{27} \zeta_3 \right) N_f + \frac{19552}{729} N_f^2\,.
\end{eqnarray}

{\underline{Two-derivative vector/axial}}~\cite{Retey:2000nq,Gracey:2006zr,Gockeler:2010yr}
\begin{eqnarray}
  \gamma_0 & = & \frac{100}{9} \,, \\
  \gamma_1^{\overline{\rm MS}} & = & \frac{34450}{243} - \frac{830}{812}\,N_f \,, \\
  \gamma_1^{\mbox{\scriptsize RI$^{\prime}$}} & = & \frac{76822}{243} - \frac{562}{27} N_f \,, \\
  \gamma_2^{\overline{\rm MS}} & = & \frac{64486199}{26244} + \frac{2200}{81}\zeta_3  - \left( \frac{469910}{2187}  + \frac{4000}{27} \zeta_3 \right) N_f - \frac{2569}{729} N_f^2\,,\\
  \gamma_2^{\mbox{\scriptsize RI$^{\prime}$}} & = & \frac{1889349409}{131220} - \frac{744568}{408} \zeta_3 - \left(\frac{20589053}{10935} + \frac{4736}{135} \zeta_3 \right) N_f + \frac{34330}{729} N_f^2\,.
\end{eqnarray}

\section{Correlation coefficients of fit parameters}
\label{sec:appB}

The chiral fit for $\langle x\rangle$ gives the following best fit
parameters
\begin{center}
  \begin{tabular*}{.3\textwidth}{@{\extracolsep{\fill}}lr}
    \hline\hline
    $c_0$ & 0.199(5) \\
    $c_1$ & 0.0083(5) \\
    $c_a$ & 0.92(20) \\
    $P_{Z_\mathrm{vD}}(\beta=1.90)$ & 1.033(9) \\
    $P_{Z_\mathrm{vD}}(\beta=1.95)$ & 1.053(7) \\
    $P_{Z_\mathrm{vD}}(\beta=2.10)$ & 1.100(6) \\
    $P_r(\beta=1.90)$ & 5.32(8) \\
    $P_r(\beta=1.95)$ & 5.76(6) \\
    $P_r(\beta=2.10)$ & 7.60(8) \\
    \hline\hline
  \end{tabular*}
\end{center}
with correlation coefficients in the same order as above:
\begin{equation*}
  \begin{matrix}
  1.0 & -0.37 & -0.84 & -0.49 & -0.32 & 0.53 & -0.24 & 0.03 & 0.09\\
      & 1.0   & -0.10 & 0.10 & 0.13 & -0.03 & 0.05 & -0.004 & 0.01\\
      &       &  1.0  & 0.68 & 0.49 & -0.43 & 0.28 & 0.02 & -0.06\\
      &       &       & 1.0  & 0.68 & -0.04 & -0.07 & 0.09 & 0.03\\
      &       &       &      & 1.0  & 0.07  & 0.22 & -0.14 & 0.07\\
      &       &       &      &      & 1.0   & -0.08 & 0.07 & -0.04\\
      &       &       &      &      &       & 1.0   & 0.01 & 0.03\\
      &       &       &      &      &       &       & 1.0 & 0.03\\
  \end{matrix}
\end{equation*}
The chiral fit for $\langle x^2\rangle$ gives the following best fit
parameters
\begin{center}
  \begin{tabular*}{.3\textwidth}{@{\extracolsep{\fill}}lr}
    \hline\hline
    $b_0$ & 0.16(2) \\
    $b_1$ & 0.005(2) \\
    $b_a$ & -1.6(7) \\
    $P_{Z_\mathrm{vDD}}(\beta=1.90)$ & 1.31(2) \\
    $P_{Z_\mathrm{vDD}}(\beta=1.95)$ & 1.35(2) \\
    $P_{Z_\mathrm{vDD}}(\beta=2.10)$ & 1.41(2) \\
    $P_r(\beta=1.90)$ & 5.30(7) \\
    $P_r(\beta=1.95)$ & 5.78(6) \\
    $P_r(\beta=2.10)$ & 7.60(8) \\
    \hline\hline
  \end{tabular*}
\end{center}
with correlation coefficients
 in the same order as above:
\begin{equation*}
  \begin{matrix}
  1.0 & -0.38 & -0.73 & -0.14 & 0.06 & 0.16 & 0.04 & 0.06 & -0.05\\
      & 1.0   & -0.33 & -0.04 & 0.05 & 0.01 & -0.03 & 0.01 & 0.004\\
      &       &  1.0  & 0.22 & -0.07 & -0.16 & -0.07 & -0.09 & 0.04 \\
      &       &       & 1.0  & 0.03 & -0.05 & 0.06 & -0.01 & 0.03\\
      &       &       &      & 1.0  & 0.01 & -0.02 & 0.05 & -0.03\\
      &       &       &      &      & 1.0   & 0.02 & 0.003 & -0.01\\
      &       &       &      &      &       & 1.0   & 0.01 & 0.04\\
      &       &       &      &      &       &       & 1.0 & 0.01\\
  \end{matrix}
\end{equation*}